\documentclass[a4paper,twoside]{article}

\usepackage{epsfig}
\usepackage{breakurl}

\usepackage{subcaption}
\usepackage{calc}
\usepackage{amssymb}
\usepackage{amstext}
\usepackage{amsmath}
\usepackage{amsthm}
\usepackage{multicol}
\usepackage{pslatex}
\usepackage{apacite}
\usepackage{multirow}
\usepackage{algorithm2e}
\usepackage[bottom]{footmisc}
\usepackage{booktabs}
\usepackage{array}
\usepackage{SCITEPRESS} 

\begin{document}

\title{Optimizing Service Placement in Edge-to-Cloud AR/VR Systems using a Multi-Objective Genetic Algorithm}

\author{\authorname{Mohammadsadeq G. Herabad\sup{1}, Javid Taheri\sup{1,2}, Bestoun S. Ahmed\sup{1,3} and Calin Curescu\sup{4}}
\affiliation{\sup{1}Department of Mathematics and Computer Science , Karlstad University, Karlstad, Sweden}
\affiliation{\sup{2}School of Electronics, Electrical Engineering and Computer Science, Queen's University Belfast, Belfast, UK
}
\affiliation{\sup{3}Department of Computer Science, Faculty of Electrical Engineering, Czech Technical University, Prague, Czech Republic
}
\affiliation{\sup{4}Ericsson AB, Stockholm, Sweden}
\email{\{mohammadsadeq.garshasbi.herabad, bestoun, javid.taheri\}@kau.se, j.taheri@qub.ac.uk, calin.curescu@ericsson.com}
}


\abstract{Augmented Reality (AR) and Virtual Reality (VR) systems involve computationally intensive image processing algorithms that can burden end-devices with limited resources, leading to poor performance in providing low latency services. Edge-to-cloud computing overcomes the limitations of end-devices by offloading their computations to nearby edge devices or remote cloud servers. Although this proves to be sufficient for many applications, optimal placement of latency sensitive AR/VR services in edge-to-cloud infrastructures (to provide desirable service response times and reliability) remain a formidable challenging. To address this challenge, this paper develops a Multi-Objective Genetic Algorithm (MOGA) to optimize the placement of AR/VR-based services in multi-tier edge-to-cloud environments. The primary objective of the proposed MOGA is to minimize the response time of all running services, while maximizing the reliability of the underlying system from both software and hardware perspectives. To evaluate its performance, we mathematically modeled all components and developed a tailor-made simulator to assess its effectiveness on various scales. MOGA was compared with several heuristics to prove that intuitive solutions, which are usually assumed sufficient, are not efficient enough for the stated problem. The experimental results indicated that MOGA can significantly reduce the response time of deployed services by an average of 67\% on different scales, compared to other heuristic methods. MOGA also ensures reliability of the 97\% infrastructure (hardware) and 95\% services (software).}

\onecolumn \maketitle \normalsize \setcounter{footnote}{0} \vfill

\section{\uppercase{Introduction}}
\label{sec:introduction}

Augmented Reality (AR) and Virtual Reality (VR) technologies revolutionise our interaction with the digital and physical world. Integration of edge and cloud computing has led to significant progress in AR and VR systems, allowing low latency data processing, which is a crucial consideration in almost all AR/VR systems \cite{elawady2020mixed}. 

Edge computing pushes data processing closer to its source, typically at the network edge, rather than relying on a centralized cloud server. On the network's edge, necessary computational tasks can be executed on edge devices and/or nearby edge servers. While edge computing handles tasks requiring low latency, cloud computing can manage resource-intensive operations through powerful computational servers in data centers \cite{ji2023towards}. Due to their complementary nature, integrating edge and cloud computing with AR/VR systems can address critical latency requirements, while enhancing AR/VR experiences by providing robust computational resources at both layers.

The complexities of real-time multimedia within AR/VR systems demand more careful consideration of how computational tasks should be distributed for execution to achieve minimum latency. This is an important concern to prevent disruption or delay in visual or interactive elements \cite{siriwardhana2021survey}. Furthermore, infrastructure and service-related reliability play an important role in providing a seamless and effective user experience \cite{dong2019reliability, khaleel2022multi}. The hardware components of the edge-to-cloud infrastructure, from the input devices to the computing nodes and the output devices, must operate uninterruptedly to provide a smooth experience. Software running AR/VR services must also be highly reliable to prevent errors and corruptions that could disrupt the user experience. The system becomes even more complex due to heterogeneous devices and computing servers coupled with various communication capacities \cite{liu2023intelligent}. 

To address the above challenges, adopting an optimization approach becomes essential to make optimal policies for placing AR/VR services on heterogeneous resources throughout the system. The optimization approach must determine decisions considering various factors including, but not limited to, memory and computing capacity of the nodes, communication characteristics between the nodes, as well as diverse service/resource (e.g., CPU, memory, disk, and network) requirements and characteristics related to required/running services. Therefore, making optimal decisions about the placement of services across available resources in the edge-to-cloud continuum can be extremely challenging and, therefore, identified (proved to be) as an NP-Hard problem \cite{raju2023delay, liu2023joint, huangpeng2024distributed}.

Although the complexity of edge-to-cloud AR/VR systems can be studied for various applications and use cases, this paper, specifically, focuses on an AR/VR-assisted remote repair/maintenance use case in industrial sectors, such as those studied in \cite{fang2020augmented}. In our use case, developed as part of a collaborative project with Ericsson, when a malfunctioning device, within an industrial application, requires repair and no expert is available on-site, a nearby technician uses an AR/VR application on their device to connect with a remote expert. The technician shares enhanced videos of the malfunctioning device with the remote expert to solicit their help for identifying/troubleshooting the instrument, and/or developing a repair plan accordingly. In this specific use case, the latency of the AR/VR application plays a vital role in its functionality as a repairing tool, and thus the optimal placement of AR/VR services for efficient offloading of the computations over the network to meet service requirements (e.g., service level agreements) is essential. 

This paper models the problem from a system design perspective and develops a Multi-Objective Genetic Algorithm (MOGA) to simultaneously address latency- and reliability-related concerns. The paper also demonstrates that heuristic algorithms are insufficient for addressing the complexities of service placement in edge-to-cloud AR/VR systems, thus justifying the use of meta-heuristic methods to achieve optimal or near-optimal solutions. We use MOGA because GA-based algorithms have been proven to be highly adaptable meta-heuristic algorithms that have been successfully applied to solve a diverse range of optimization problems. The capability of GA to handle large-scale problems and its inherent structure for parallelism also makes it a preferable choice for addressing the stated problem in this paper. Our contributions in this paper can be highlighted as follows:

\begin{itemize}
  \item We modelled a multi-tier edge-to-cloud infrastructure with the focus on AR/VR systems that need to support service components with multiple versions (e.g., different codecs). 
  \item We proposed a multi-objective GA-based service placement approach to simultaneously minimize both service response time and maximize infrastructure/service reliability.
  \item We proposed a grid-based fine-tuning procedure to identify optimal configurations (population size, crossover rate, etc.) for our MOGA, aiming to achieve a balance between the fitness and runtime of MOGA.
  \item We developed a tailor-made simulator to evaluate the performance of MOGA as compared with other heuristic algorithms. We use multiple metrics and explain how/why MOGA provides an optimal service placement solution in edge-to-cloud AR/VR systems.
\end{itemize}

The remainder of this paper is structured as follows. Section \ref{sec:RelatedWork} provides the related work. Section \ref{sec:SystemModel} formulates the infrastructure and the service model. Section \ref{sec:ObjectiveFunction} defines the objective function. In Section \ref{sec:ProposedMOGA}, our solution is introduced. Section \ref{sec:Evaluation} outlines the experimental setup for simulations and evaluations, providing an analysis of the results obtained through the simulations. Finally, Section \ref{sec:Conclusion} presents the paper conclusion.

\section{\uppercase{Related Work}} 
\label{sec:RelatedWork}
Over the past few years, many approaches were designed to use different techniques for offloading video streams from end-devices to the near edge devices or clouds to reduce service delays and service response time in edge-to-cloud environments \cite{ren2021adaptive, acheampong2023parallel, chen2022unsupervised, huang2021proactive}. Specifically, the authors of \cite{cozzolino2022nimbus} proposed an approach called Nimbus, which is a multi-objective solution for task allocation in edge-to-cloud environments. The primary goal of Nimbus is to minimize the latency of AR applications by offloading computational tasks to edge or cloud servers. This work, however, focuses on simple AR tasks without paying sufficient attention to the complications associated with multi-version AR tasks and procedures. Additionally, authors of \cite{yeganeh2023novel} devised an approach that uses Q-learning to reduce task execution time and mitigate energy consumption on end-devices by offloading computational tasks to edge-nodes. Through a series of experiments, the authors claimed that their approach provides a better solution, as compared with similar algorithms that were implemented in the article in 90\% of the cases. The paper addresses the problem from a general perspective, without explicitly focusing on the complexity of video streaming.

Authors of \cite{mahjoubi2022ehga} formulated service placement as a Mixed-Integer Linear Programming (MILP) problem in edge-to-cloud computing. Subsequently, a single-objective genetic algorithm \cite{mahjoubi2022ehga} and a simulated annealing algorithm \cite{mahjoubi2022efficient} were used to solve the problem. These works focused on single-objective algorithms, aiming to minimize the runtime of the system without addressing other metrics, such as system reliability, in terms of software and hardware aspects. The authors of \cite{de2023bee} also introduced an approach based on Bee Colony optimization to reduce the execution time of the application for computational tasks by offloading them to the edge. The extensive experiments described in the paper supported the claim that their technique can reduce the execution time by 56\% when compared with other heuristics. This work neglected to take into account the computational capacity of the cloud, particularly in scenarios with substantial computational demands. This is because the algorithm primarily focuses on the resource conditions of end devices and edge servers to make offloading decisions. The authors of \cite{fan2022collaborative} presented a collaborative approach based on the Lyapunov optimization technique for the placement of services within edge-to-cloud systems. Although this work demonstrated the effectiveness of the proposed algorithm in various scenarios to minimize overall task processing delay and ensure long-term task queue stability, it focused on single-objective scenarios where the main objective is to reduce the system delay. Similarly, the authors of \cite{wang2023microservice} formulated service placement as a linear integer programming problem and proposed a polynomial-time method to make decisions collaboratively among edge nodes and handle service heterogeneity in the system. This research focuses on service placement in small-scale scenarios, neglecting to explicitly consider aspects related to software and hardware reliability.

Besides approaches for single-objective problems, there were also several works focusing on multi-objective ones, such as \cite{abedi2022dynamic, li2022multiobjective, lavanya2020multi, madni2019multi}. Despite the significant body of research investigating service placement and offloading strategies within edge-to-cloud computing environments, a limited portion of these studies consider the complexities of edge-to-cloud AR/VR systems. In fact, many studies adopted a broad perspective, failing to consider the unique challenges and requirements posed by the nature of AR/VR workloads. As a result, there is a notable gap in the literature on approaches that address the specific characteristics of AR/VR systems in terms of service response time, hardware reliability, and service reliability, specifically in cases where multiple versions of AR/VR service components (co-)exits in a platform. This gap deserves further investigation and exploration in this domain, and motivated us to design algorithms in this article.

\section{\uppercase{System Model}}
\label{sec:SystemModel}
This section presents mathematical models for a multi-tier infrastructure that hosts AR/VR-based services.

\subsection{User and helper model}
We consider a use case where `users' can connect with remote `helpers' using AR/VR applications. This connection allows users to consult with remote helpers (called user-helper pairs) to perform repair or maintenance tasks on industrial sites. 

Both users and helpers have their own devices, denoted by $U = \lbrace u_{1}, u_{2}, ..., u_{n}, ..., u_{N}\rbrace$ and $H = \lbrace h_{1}, h_{2}, ..., h_{m}, ..., h_{M}\rbrace$, respectively. Resources for users and helpers are modeled as nodes with computation capacity (CC), memory capacity (MM) and disk capacity (DC), as well as other characteristics such as an Operating System (OS) and a reliability score (RS). These characteristics are indicated by $\mathit{u_n} = \langle CC_n, MC_n, DC_n, OS_n, RS_n\rangle$ and $\mathit{h_m} = \langle CC_m, MC_m, DC_m, OS_m, RS_m\rangle$; we define $U^{ch}$ and $H^{ch}$ (Equations \eqref{eq:1} and \eqref{eq:2}) to represent the characteristics of all users and helpers, respectively. We also define $P = \lbrace p_1, p_2, ..., p_i, ... p_I\rbrace$ to show the set of pairs, where $p_i = \langle u_n, h_m\rangle$ represents the $i^{th}$ pair. Table \ref{table:userhelpernotations} describes the notation related to the user and helper nodes.

\begin{equation}\label{eq:1}
U^{ch} = \begin{bmatrix}CC_{1} & MC_{1} & DC_{1} & OS_{1} & RS_{1}
                \\CC_{2} & MC_{2} & DC_{2} & OS_{2} & RS_{2}
                \\ \vdots & \vdots & \vdots & \vdots & \vdots
                \\ CC_{N} & MC_{N} & DC_{N} & OS_{N} & RS_{N}\end{bmatrix}
\end{equation}

\begin{equation}\label{eq:2}
H^{ch} = \begin{bmatrix}CC_{1} & MC_{1} & DC_{1} & OS_{1} & RS_{1}
                \\CC_{2} & MC_{2} & DC_{2} & OS_{2} & RS_{2}
                \\ \vdots & \vdots & \vdots & \vdots & \vdots
                \\ CC_{M} & MC_{M} & DC_{M} & OS_{M} & RS_{M}\end{bmatrix}
\end{equation}

\begin{table}
    \caption{Notations related to the user and helper nodes}
    \label{table:userhelpernotations}
    \renewcommand{\arraystretch}{1}
    \small
    \begin{tabular}{m{1cm} m{5.5cm}}
        \toprule
        Notation & Description \\
        \midrule
        \midrule
        $U$ & The set of user nodes \\
        \midrule
        $H$ & The set of helper nodes \\
        \midrule
        $u_{n}$ & The $n^{th}$ user node \\
        \midrule
        $h_{m}$ & The $m^{th}$ helper node \\
        \midrule
        $U^{ch}$ & The characteristics of all user nodes \\
        \midrule
        $H^{ch}$ & The characteristics of all helper nodes \\
        \midrule
        $P$ & The set of pairs \\
        \midrule
        $p_i$ & The $i^{th} pair$ \\
        \midrule
        $N$ & The total number of user nodes \\
        \midrule
        $M$ & The total number of helper nodes \\
        \midrule
        $I$ & The total number of pairs \\
        \midrule
        $CC$ & Computation capacity (MIPS) \\
        \midrule
        $MC$ & Memory capacity (GB) \\
        \midrule
        $DC$ & Disk capacity (GB) \\
        \midrule
        $RS$ & Reliability score \\
        \midrule
        $OS$ & Operating system \\
        \bottomrule
    \end{tabular}
\end{table}

\subsection{Service model}
The system hosts multiple AR/VR services, $S = \lbrace S_{1}, S_{2}, ..., S_{x}, ..., S_{X} \rbrace$, where $S_{x}$ denotes the service running through the $x^{th}$ user-helper pair.
Each service, consists of $Y$ service components, where $SC_y^x$ represents service component $y$ of $x^{th}$ user-helper pair, $S_{x} = \lbrace SC_1^x, SC_2^x, ..., SC_y^x, ..., SC_Y^x\rbrace$. Similarly, every service component (in a service) has $V$ versions, where $SC_{y,v}^x$ represents version $v$ of the service component $y$ of the service belonging to pair $x$, $SC_y^x = \lbrace SC_{y,1}^x, SC_{y,2}^x, ..., SC_{y,v}^x, ..., SC_{y,V}^x\rbrace$. Each version of a service component has specific requirements (demands) regarding computational power, memory capacity, and disk space to ensure successful execution; it also produces/transfers data (with specific size) to other dependent service components. Furthermore, each version is associated with various providers, including external providers such as AWS and Azure, and internal providers such as hosted Kubernetes clusters. There are also attributes for service components, such as Transcode, Codec Type, and a reliability score to indicate how they work and how often they fail. We define ${SC_{y^{ch}}^x}$ to represent all the resource requirements and characteristics of all versions of a service component (Table \ref{table:servicenotations} shows the notation of the service model.)

\begin{equation}\label{eq:3}
    \scriptsize
    {SC_{y^{ch}}^x} = \begin{bmatrix}
        CR_{1} & MR_{1} & DR_{1} & DS_{1} & PR_{1} & TC_{1}& CT_{1} & {RS}_{1} \\
        CR_{2} & MR_{2} & DR_{2} & DS_{2} & PR_{2} & TC_{2}& CT_{2} & {RS}_{2} \\
        \vdots & \vdots & \vdots & \vdots & \vdots & \vdots & \vdots & \vdots\\
        CR_{V} & MR_{V} & DR_{V} & DS_{V} & PR_{V} & TC_{K}& CT_{V} & {RS}_{V}
    \end{bmatrix}
\end{equation}

Where ${S_{x^{ch}}}$ denotes the resource requirements and characteristics of all service components belonging to the $x^{th}$ user-helper pair.

\begin{equation}\label{eq:4}
    \footnotesize
    {S_{x^{ch}}} = \begin{bmatrix}
        SC_{1^{ch}}^x & SC_{2^{ch}}^x & ... & SC_{y^{ch}}^x &... & SC_{Y^{ch}}^x
    \end{bmatrix}
\end{equation}

\begin{figure}
    \centering
    \includegraphics[width=1\linewidth]{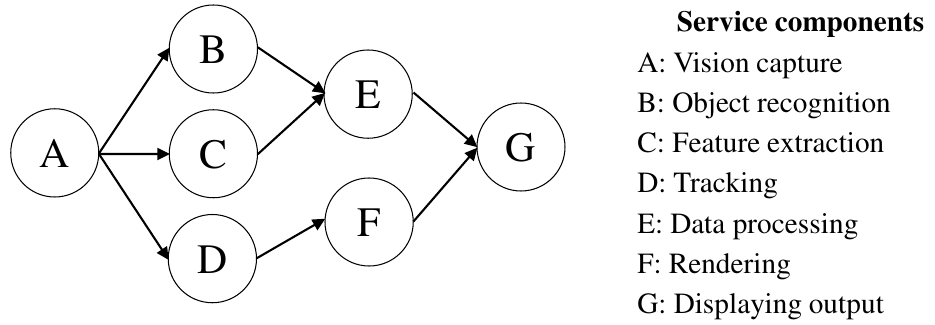}
        \captionsetup{
        justification=centering,
        font={small,stretch=0.9}
    }
    \caption{An example DAG for an AR/VR application}
    \label{fig:dag}
\end{figure}

The service components of each service are interdependent. Therefore, each service can be represented by a Directed Acyclic Graph (DAG), where the vertices in the DAG represent the service components, and the links between the vertices represent the dependencies between the service components. Figure \ref{fig:dag} shows an example of a DAG that represents relationships between the service components of an AR/VR application. In this paper, an upper binary triangular matrix is used to model the DAG, where a value of `1' in the matrix implies a connection between the corresponding service components.

\begin{table}
    \caption{Notations related to service model}
    \label{table:servicenotations}
    \renewcommand{\arraystretch}{0.8}
    \small
    \begin{tabular}{m{1cm} m{5.5cm}}
        \toprule
        Notation & Description \\
        \midrule
        \midrule
        $S$ & The set of services \\
        \midrule
        $S_{x}$ & The service belongs to the $x^{th}$ user-helper pair \\
        \midrule
        $S_{x^{ch}}$ & The characteristics of all service components of $S_x$ \\
        \midrule
        $SC_y^x$ & The $y^{th}$ service component of $S_{x}$  \\
        \midrule
        $SC_{y,v}^x$ & The $v^{th}$ version of the $y^{th}$ service component of $S_{x}$ \\
        \midrule
        $SC_{y^{ch}}^x$ & The characteristics of all versions of $SC_y^x$ \\
        \midrule
        $X$ & The total number of services (X = I) \\
        \midrule
        $Y$ & The total number of service components of each service \\
        \midrule
        $V$ & The total number of service component versions  \\
        \midrule
        $CR$ & Computation requirement of a service component (MIPS) \\
        \midrule
        $MR$ & Memory requirement of a service component (GB) \\
        \midrule
        $DR$ & Disk requirement of a service component (GB) \\
        \midrule
        $DS$ & The data size that is required to transfer (Mb) \\
        \midrule
        $PR$ & Service component provider \\
        \midrule
        $TC$ & Service component transcode \\
        \midrule
        $CT$ & Service component codec type \\
        \bottomrule
    \end{tabular}
\end{table}

\subsection{Infrastructure and network model}
\begin{figure}
    \centering
    \includegraphics[width=0.9\linewidth]{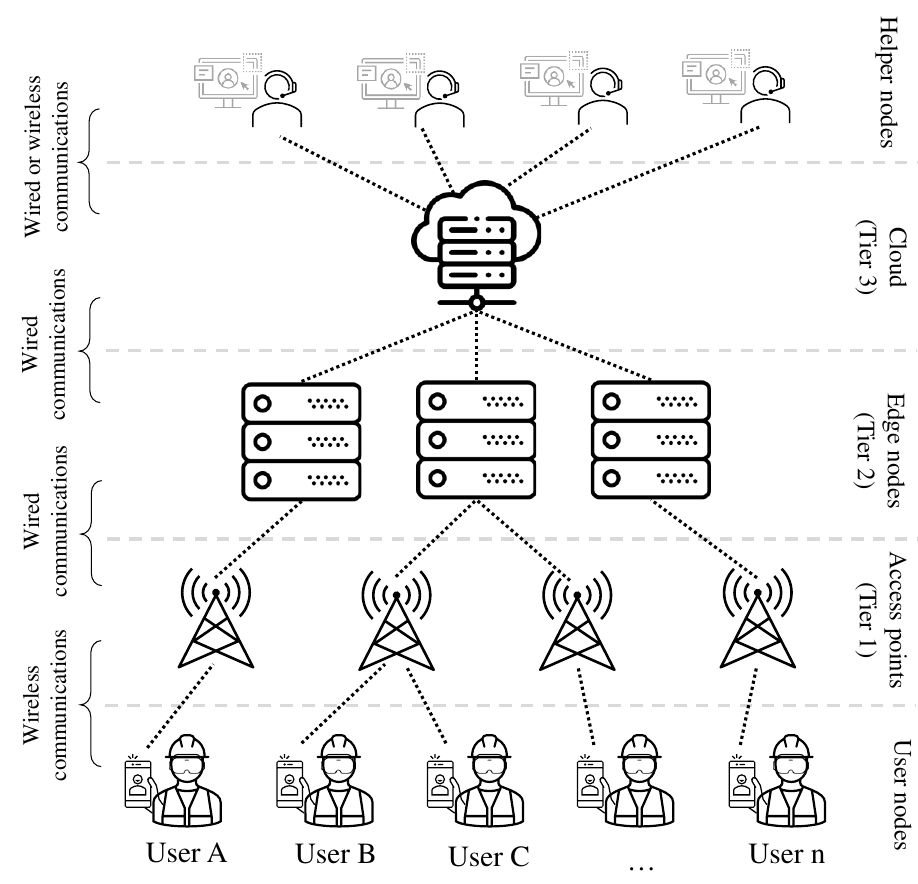}
    \captionsetup{
        justification=centering,
        font={small,stretch=0.9}
    }
    \caption{Multi-tier edge-to-cloud infrastructure}
    \label{fig:infrastructure}
\end{figure}

We employ a three-tier infrastructure, including access points (AP), edge, and cloud tiers. Tier-1 is defined as a group of APs with specific computing capabilities. These APs establish direct wireless connections with users, acting as network entry points. Tier-2 comprises a collection of computing nodes positioned close to (in terms of network accessibility and speed) backbone routers in proximity to Tier-1. Nodes within Tier-2 have enhanced computational and storage capacities. Tier-3 contains the cloud in our system, providing the highest computational power and storage capacity among all tiers. The increase in computing capability in the upper tiers comes with a rise in data transmission latency. We assume that the helpers are connected (accessible) through the cloud (because they are located in remote places). Figure \ref{fig:infrastructure} illustrates the entities in our system.

The computing nodes in the different tiers have their own device characteristics, denoted as $\mathit{CN} = \lbrace CN_1, CN_2, ..., CN_k, ..., CN_K \rbrace$, where $\mathit{CN_k} = \langle CC_k, MC_k, DC_k, OS_k, RS_k \rangle$. Moving to the upper tiers, the memory and computational capacity of computing nodes increase with respect to CPUs, GPUs, and RAM, while their network accessibility decreases (i.e., communicating with them incurs more delay). We define $CN^{ch}$ to represent the characteristics of the computing nodes within the infrastructure.

\begin{equation}\label{eq:5}
   CN^{ch} = \begin{bmatrix}
        CC_{1} & MC_{1} & DC_{1} & OS_{1} & RS_{1} \\
        CC_{2} & MC_{2} & DC_{2} & OS_{2} & RS_{2} \\
        \vdots & \vdots & \vdots & \vdots & \vdots \\
        CC_{K} & MC_{K} & DC_{K} & OS_{K} & RS_{K}
    \end{bmatrix}
\end{equation}

To model the network communications among computing nodes within the infrastructure and between users, helpers, and computing nodes, we define $CN^{L}$ (Equation \ref{eq:6}) to represent the available bandwidth (BW) and transmission delays (LD) on each communication link. For example, $BW_{2,1}$ shows the available bandwidth between computing nodes 2 and 1, similarly, $LD_{2,1}$ shows observed delay between computing nodes 2 and 1. In Equation \ref{eq:6} that represents $CN^{L}$, we assume that rows and columns from 1 to $K$ are related to computing nodes, columns from $K+1$ to $K+N$ are associated with user nodes, and columns from $K+N+1$ to $K+N+M$ are related to helper nodes. Here, each element represents available bandwidth and observed delay (e.g., half of the ICMP round-trip time between two entities) among users, helpers, and computing nodes. We use ICMP round-trip time to consider all potential delays between two nodes, including those resulting from routing processes. Since the users are not connected to each other and the helpers also are not connected to each other, the number of rows in the matrix is equal to $K$. Table \ref{table:infrastructure} describes the notation related to the infrastructure and the network model.

\begin{equation}\label{eq:6}
    \scriptsize
{CN^{L}} = \begin{bmatrix} \langle0,0\rangle & \dotsm & \langle BW_{1,K+N+M},LD_{1,K+N+M}\rangle\\

                  \langle BW_{2,1},LD_{2,1} \rangle & \dotsm & \langle BW_{2,K+N+M},LD_{2,K+N+M}\rangle\\
                  \vdots & \ddots& \vdots\\
                  \langle BW_{K,1},LD_{K,1} \rangle & \dotsm& \langle BW_{K,K+N+M},LD_{K,K+N+M}\rangle\\
            \end{bmatrix}
\end{equation}

\begin{table}
    \caption{Notations of the infrastructure and network model}
    \label{table:infrastructure}
    \renewcommand{\arraystretch}{1}
    \small
    \begin{tabular}{m{1cm} m{5.5cm}}
        \toprule
        Notation & Description \\
        \midrule
        \midrule
        $CN$ & The set of computing nodes in the infrastructure \\
        \midrule
        $CN^{ch}$ & The resource characteristics of the computing nodes \\
        \midrule
        $CN^{L}$ & The characteristics of communication links among users, helpers, and computing nodes \\
        \midrule
        $BW_{K,1}$ & The available bandwidth between node $K$ and node $1$ (Mbps) \\
        \midrule
        $LD_{K,1}$ & The link delay between node $K$ and node $1$ (ms) \\
        \midrule
        $K$ & The number of total computing nodes \\
        \bottomrule
    \end{tabular}
\end{table}

\section{\uppercase{Optimization Model}}
\label{sec:ObjectiveFunction}
We adopt a centralized service placement approach, where service placement algorithms operate in the cloud to decide on the placement of AR/VR services across the entire infrastructure. We assume that data related to services and infrastructure characteristics are available before making service placement decisions. To formulate an objective function for optimal decision-making and service placement, various factors are considered, and discussed in the following subsections.

\subsection{Data transmission delay model}
The data transmission delay depends on both the network bandwidth and the data size of the service components. In addition, other factors, such as jitter, can affect the transmission time of data. Therefore, we consider the links' bandwidth, the data size of the service components, and the Round Trip Time (RTT), which can be measures proactively using ICMP messages, of the links to calculate transmission delays \cite{cozzolino2022nimbus}. Equation \eqref{eq:7} calculates the delay of data transmission, where $DS_v$ shows the required data size transferred by version $v$ of the service component and $BW$ shows the bandwidth of the transmission link. Also, we divide the RTT by 2 to account for the one-way delay of the link.

\begin{equation}\label{eq:7}
TD(SC_{y,v}^x) = \frac{DS_v}{BW} + \frac{RTT}{2}
\end{equation}

If a service component needs to transfer data to multiple service components in different places, equation \eqref{eq:7} is calculated separately for each. The notations related to the transmission delay model are presented in Table \ref{table:transmission}.

\begin{table}
    \caption{Notations of the transmission delay model}
    \label{table:transmission}
    \renewcommand{\arraystretch}{1}
    \small
    \begin{tabular}{m{1.2cm} m{5.3cm}}
        \toprule
        Notation & Description \\
        \midrule
        \midrule
        $DS_v$ & The data size transferred by $v^{th}$ version of the service component \\
        \midrule
        $TD({SC_{y,v}^x})$ & The transmission delay of $SC_{y,v}^x$ \\
        \midrule
        $BW$ & The link bandwidth \\
        \midrule
        $RTT$ & Round trip time of the link \\
        \bottomrule
    \end{tabular}
\end{table}

\subsection{Execution time model}
The execution time of a service component is closely related to the computational capacity of the computing node and the computational requirements of the service component. Consequently, the execution time of a service component is calculated using Equation \eqref{eq:8}. $CR_v$ reflects the total number of instructions required to execute a version $v$ of the service component, and $CC_k$ reflects the computational capacity of the computing node $k$. We assume the computing nodes support multi-threading and can execute the service components concurrently. However, a waiting time is considered in execution time, denoted as $W_v$. This waiting time occurs when a service component has been placed on the computing node but has not started its execution on the node yet. The notations related to the execution time model are described in Table \ref{table:execution}.

\begin{equation}\label{eq:8}
ET({SC_{y,v}^x}) = \frac{CR_v}{CC_k} + W_v
\end{equation}

\begin{table}
    \caption{Notations related to the the execution time model}
    \label{table:execution}
    \renewcommand{\arraystretch}{1}
    \small
    \begin{tabular}{m{1cm} m{5.5cm}}
        \toprule
        Notation & Description \\
        \midrule
        \midrule
        $CC_k$ & The computation capacity of computing node $k$ \\
        \midrule
        $CR_v$ & The required computation of the $v^{th}$ version of $SC_{y}^x$ \\
        \midrule
        $W_v$ & The waiting time of the $v^{th}$ version of $SC_{y}^x$ \\
        \midrule
        $ET({SC_{y,v}^x})$ & The execution time of $SC_{y,v}^x$ \\
        \bottomrule
    \end{tabular}
\end{table}

\subsection{Response time model}
To calculate the response time of a service component, various delays are considered: provider delays and coding delays, denoted as $PD({SC_{y,v}^x})$ and $CD({SC_{y,v}^x})$, respectively. Provider delays are assumed to have (almost) constant values and can be estimated through passive measurements on the network. Coding delays reflect the execution time of specific versions of encoding and decoding algorithms; they also assumed constant and can be obtained from the algorithm providers and/or through multiple runs on a sample video feed.

The response time for a service component is calculated using Equation \eqref{eq:9}. Similarly, the total response time for a service and all services are obtained by \eqref{eq:10} and \eqref{eq:11}, respectively. Table \ref{table:responsetime} describes notations of the response time model.

\begin{multline}\label{eq:9}
\small
RT({SC_{y,v}^x}) = TD({SC_{y,v}^x}) + ET({SC_{y,v}^x}) \\ + PD({SC_{y,v}^x})
 + CD({SC_{y,v}^x})
\end{multline}

\begin{equation}\label{eq:10}
RT({S_{x}}) = \displaystyle\sum_{y = 0, v \in V}^Y RT({SC_{y,v}^x})
\end{equation}

\begin{equation}\label{eq:11}
RT({S}) = \displaystyle\sum_{x = 0}^X RT({S_x})
\end{equation}

\begin{table}
    \caption{Notations related to the response time model}
    \label{table:responsetime}
    \renewcommand{\arraystretch}{1}
    \small
    \begin{tabular}{m{1.2cm} m{5.3cm}}
        \toprule
        Notation & Description \\
        \midrule
        \midrule
        $RT({SC_{y,v}^x})$ & The response time of ${SC_{y,v}^x}$\\
        \midrule
        $RT({S_{x}})$ & The response time of $S_{x}$ \\
        \midrule
        $RT({S})$ & The response time of $S$ \\
        \midrule
        $PD({SC_{y,v}^x})$ & The provider delay of ${SC_{y,v}^x}$ \\
        \midrule
        $CD({SC_{y,v}^x})$ & The encoding and decoding delay of ${SC_{y,v}^x}$ \\
        \bottomrule
    \end{tabular}
\end{table}

\subsection{Reliability model}
The reliability of both the nodes and the service components is measured by their historical performance. In particular, researchers of \cite{moghaddas2016reliability, amini2022new, liu2015schedule} have introduced methods to measure their reliability scores. Taking into account the interdependent relationships between the components of the service, the overall reliability of the service (software) is calculated using Equation \eqref{eq:12} \cite{maciel2021survey} to determine the probability of successfully completing a service.

\begin{equation}\label{eq:12}
RS({S_{x}}) = \displaystyle\prod_{y = 0, v\in V}^Y {RS_{SC_{y,v}^{x}}}
\end{equation}

The average reliability of all services is calculated using Equation \eqref{eq:13}.

\begin{equation}\label{eq:13}
RS({S}) = \frac{\sum_{x=0}^X RS({S_{x}})}{X}
\end{equation}

Taking into account the independent relationship among the computing nodes in the infrastructure, the total reliability of the computing nodes is represented by Equation \eqref{eq:14} \cite{maciel2021survey}, where $RS_k$ measures the reliability score of a computing node $k$.
\begin{multline}\label{eq:14}
RS(CN) = 1 - (1 - RS_1) \times \dotsm \times \\
(1-RS_k) \times \dotsm \times (1 - RS_K)
\end{multline}

In a user-helper pair, because the failure of one part (i.e., user node, helper node or computing nodes) can adversely affect other parts, the overall hardware reliability of the pairs is determined using Equation \eqref{eq:15} \cite{maciel2021survey}. Table \ref{table:reliability} describes the notations of the reliability model.

\begin{equation}\label{eq:15}
RS(P) = RS(CN) \cdot RS(U) \cdot RS(H)
\end{equation}

\begin{table}
    \caption{Notations related to the reliability model}
    \label{table:reliability}
    \renewcommand{\arraystretch}{1}
    \small
    \begin{tabular}{m{0.9cm} m{5.5cm}}
        \toprule
        Notation & Description \\
        \midrule
        \midrule
        $RS_{SC_{y,v}^{x}}$ & The reliability score of ${SC_{y,v}^{x}}$\\
        \midrule
        $RS_k$ & The reliability score of a computing node $k$ \\
        \midrule
        $RS({S_{x}})$ & The reliability score of $S_{x}$ \\
        \midrule
        $RS({S})$ & The average reliability of all services \\
        \midrule
        $RS(CN)$ & The reliability of computing nodes \\
        \midrule
        $RS(U)$ & The average reliability score of the users \\
        \midrule
        $RS(H)$ & The average reliability score of helpers \\
        \midrule
        $RS(P)$ & The total hardware reliability of the pairs \\
        \bottomrule
    \end{tabular}
\end{table}

\subsection{Objective function}
Our main objective is to minimize the total response time of all services, while maximizing both hardware and software reliability. Therefore, our primary function is to map service components to heterogeneous computing nodes, as formulated in Equation \eqref{eq:16}.

\begin{equation}\label{eq:16}
 Objective function: \begin{cases}min\:\:RT({S}) & \\max\:\:RS(S) & \\ max\:\:RS(P)\end{cases}
\end{equation}
\textit{subjected to:}
\begin{equation}\label{eq:17}
\displaystyle\sum_{y, v\in Y,V} SC_{y,v}^x = 1, \forall y
\end{equation}
\begin{equation}\label{eq:18}
\displaystyle\sum_{x,y,v\in X,Y,V} res(SC_{y,v}^x) \:\:< \:\:res(CN_k), \forall k
\end{equation}
\begin{equation}\label{eq:19}
SC_{y,v}^x = u_x,\:\:\:\:\forall x
\end{equation}
\begin{equation}\label{eq:20}
SC_{y,v}^x = h_x, \:\:\:\:\forall x
\end{equation}

The constraint \eqref{eq:17} shows that a service component within the system must be exclusively assigned to a single computing node (no additional copies of a service component are allowed). The constraint \eqref{eq:18} reflects that the total resources required for all service components running on the computing node ($\displaystyle\sum res(SC_{y,v}^x)$) must not exceed the available resources of the computing node ($res(CN_k)$). The constraint \eqref{eq:19} shows that each user device is restricted to execute only service components that are directly associated with that user. Similarly, the constraint \eqref{eq:20} explicitly requires helper devices to run service components that belong to that helper node and its user. Because the formalized objective function belongs to the class of problems (NP-Complete) that cannot be solved in polynomial time, we developed our approach (MOGA) to efficiently solve this problem.

\section{\uppercase{Proposed Solution (MOGA)}}
\label{sec:ProposedMOGA}
Our solution begins the optimization process by defining the objective function and generating the initial population. Each individual in the population is assessed by a fitness function. Genetic operators, including crossover and mutation, generate offspring solutions. The population is then updated by replacing individuals with new offspring using a selection operator. MOGA also applies a healing operator to individuals to heal (rather than discard) chromosomes that partially violate constraints. The algorithm continues to iterate via these steps until a termination criterion is fulfilled. Table \ref{table:moga} describes MOGA parameters.

\begin{table}
    \caption{Notations related to MOGA parameters}
    \label{table:moga}
    \renewcommand{\arraystretch}{1}
    \small
    \begin{tabular}{m{1.5cm} m{5cm}}
        \toprule
        Notation & Description \\
        \midrule
        \midrule
        $ps$ & Populations size\\
        \midrule
        $cr$ & Crossover rate \\
        \midrule
        $mr$ & Mutation rate \\
        \midrule
        $ss$ & Tournament selection size \\
        \midrule
        $it$ & Number of iterations \\
        \midrule
        $f(i)$ & Fitness of $i^{th}$ chromosome \\
        \midrule
        $w_1$, $w_2$, $w_3$ & A value between 0 and 1 where $w_1+w_2+w_3 = 1$ \\
        \bottomrule
    \end{tabular}
\end{table}

\subsection{Chromosome encoding}
\begin{figure*}
\centering
\includegraphics[width=0.9\textwidth]{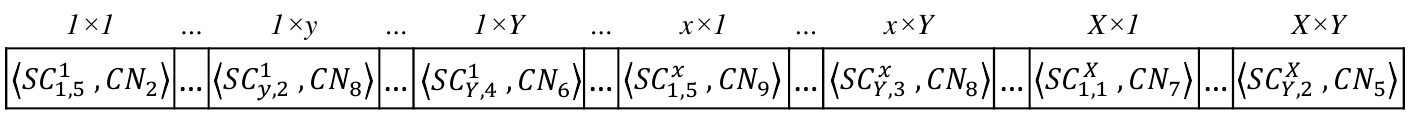}
    \captionsetup{
        justification=centering,
        font={small,stretch=0.9} 
    }
\caption{Chromosome encoding}
\label{fig:Chromosome}
\end{figure*}

Each chromosome is represented using an array of $X\times Y$ elements. Each element is considered as a gene, consisting of a tuple with two values: the first value indicates the versions of the service components and the second value is associated with the computing node running the version of the service component. Figure \ref{fig:Chromosome} represents an example chromosome, where the element ${x\times Y}^{th}$ shows the $Y^{th}$ service component of service $x$ is using the version $3$ and runs on the computing node $8$. 

To generate the initial population of chromosomes, random versions of service components are arbitrarily assigned to computing nodes. By introducing randomness in both the assignment of the versions and computing nodes, the initial population covers a broad search space, leading to the exploration of a diverse solution space.

\subsection{Fitness function}
For each chromosome, three values are calculated: response time, infrastructure reliability, and service reliability (discussed in section \ref{sec:ObjectiveFunction}). A lower response time and higher values for both infrastructure and service reliabilities indicate a superior chromosome. To this end, we transform the reliabilities' values to create a minimization fitness function. It is important to note that the reliabilities are constrained within the range of 0 to 1 ($1 \geq RS(P) > 0$, $1 \geq RS(S) > 0$).

After calculating the values, normalization is performed to ensure consistent scaling. Subsequently, a fitness function is defined using the weighted sum approach, denoted in Equation \eqref{eq:21}, where $i$ reflects the $i^{th}$ chromosome and $f(i)$ is a combined fitness function of the chromosome. Given the importance of both response time and system reliability in improving overall service quality, we consider equal weights for $RT(S)$, $RS(P)$, and $RS(S)$, where $w_1 = 0.33, w_2 = 0.33$ and $w_3 = 0.33$ (although other weights can be set to match other use cases). Because the fitness function is a minimization function, MOGA operates to minimize its value, and thus lower fitness values indicate better chromosomes.

\begin{equation}\label{eq:21}
\small
f(i) = w_1\cdot RT(S) + w_2\cdot (1-RS(P)) + w_3\cdot (1-RS(S))
\end{equation}

\subsection{Selection operator}
A tournament strategy is empirically chosen as the selection operator; Other selection procedures, such as the roulette wheel, led to the generation of more populations for convergence. In tournament selection, a number of chromosomes ($ss$) are randomly chosen and then ranked based on their fitness values. The one with the best fitness value is selected to be part of the next-generation population. These steps are repeated iteratively until the target population size ($ps$) is achieved for the next generation.

\subsection{Crossover and mutation operators}
Crossover operator can be accomplished using various techniques such as single-point crossover, multi-point crossover, and uniform crossover. In this paper, we empirically employed the single-point crossover technique, where two chromosomes (called parent pair) among the population with a crossover rate ($cr$) are selected for the crossover operation. The crossover point is randomly generated for the parent chromosomes. Genes located after the crossover point on one parent chromosome are exchanged with genes located before the crossover point in the other parent chromosome to generate two offspring. Then, the parent pair is replaced with the offspring in the population. We use the single-point crossover because it can make a trade-off between exploitation and the operator's run-time, particularly when dealing with substantial population sizes and extensive search spaces, making it a preferable choice over other crossover techniques. Other methods such as multi-point and uniform crossover also led to more complicated healing processes and consequently more convergence time.

The mutation operator randomly modifies the values of a gene, allowing the algorithm to explore a broader search space and preventing premature convergence to local optimum points. In MOGA, each gene on a chromosome is subjected to mutation with a mutation rate ($mr$). We utilize an insertion mutation technique that involves assigning random valid values to the selected gene.

\subsection{Healing operator}
After crossover and mutation operators, it is inevitable that some of the offspring may not satisfy all the constraints of the objective function. Therefore, in this paper, a healing operator is designed to guarantee the satisfaction of all constraints by modifying some genes on such chromosomes. Since the constraints \eqref{eq:17} are met during the chromosome encoding process, the healing operator is designed to satisfy the constraints \eqref{eq:18}, \eqref{eq:19}, and \eqref{eq:20}. Specifically, the healing operator examines the chromosome after applying the crossover and mutation operators. If constraint \eqref{eq:18} is not met for a specific user/helper/computing node, service components of the node is reassigned to other nodes with available computing capabilities until constraint \eqref{eq:18} is satisfied. Then it updates the chromosome based on the new assignment. Similarly, if a service component does not belong to a user-helper pair but is running on the user or helper node, it is assigned to other nodes to satisfy the constraints \ref{eq:19} and \ref{eq:20}. After the healing operator, the new generation of chromosomes is determined through the selection operator.

\section{\uppercase{Experimental setup}}
\label{sec:Evaluation}
For the evaluation, we implemented a tailor-made simulator in Node.js to simulate the entire infrastructure and services. The simulator was implemented to precisely match our problem, reflecting the desired scenarios and objectives. The simulator accepts inputs in the form of JSON objects/files where the infrastructure, the AR/VR service, and the users/helpers are defined. We designed a client-server architecture for the simulator, where the client-side sends all information of a problem instance (the infrastructure, service characteristics, etc.) to the server-side in the JSON format. Subsequently, the server employs various solvers, including MOGA, to determine the optimal solution. The results are then sent back to the client side. The simulator is containerized and, along with all problem instances in this paper, are made available for access from a GitHub \cite{simulator} address. WiKis and YAML files are also provided to facilitate deploying the prepared docker-images on Kubernetes platforms.

\begin{table*}
    \caption{Evaluation scenarios and implementation setup}
    \centering
    \label{table:scenarios}
    \renewcommand{\arraystretch}{1}
    \small
    \begin{tabular}{m{3cm} m{2cm} m{2cm} m{2cm} m{2cm}}
        \toprule
        Specifications & Small-scale & Medium-scale & Large-scale & xLarge-scale \\
        \midrule
        \midrule
        N/M & 15/8 & 50/25 &  100/50 & 250/125 \\
        \midrule
        No. of CN in Tier-1/2/3 & 10/8/2 & 30/18/4 &  75/60/8 & 150/100/15 \\
        \midrule
        X/Y & 15/5 & 50/5 &  200/5 & 250/5 \\
        \midrule
        V & 5 & 6 & 7 & 8 \\
        \midrule
        \multirow{1}{*}{CR of SC} & \multicolumn{4}{c}{[800 - 3000] (MIPS)}\\
        \midrule
        \multirow{1}{*}{MR of SC} & \multicolumn{4}{c}{[1.5 - 3.3] (GB)} \\
        \midrule
        \multirow{1}{*}{DR of SC} & \multicolumn{4}{c}{[1 - 3] (GB)} \\
        \midrule
        \multirow{1}{*}{CC of U/H} & \multicolumn{4}{c}{[500 - 2200]/[1500 - 2500] (MIPS)} \\
        \midrule
        \multirow{1}{*}{MC of U/H} & \multicolumn{4}{c}{[2 - 4]/[2 - 4] (GB)} \\
        \midrule
        \multirow{1}{*}{DC of U/H} & \multicolumn{4}{c}{[4 - 8]/[4 - 8] (GB)} \\
        \midrule
        \multirow{1}{*}{CC of CN in Tier-1/2/3} & \multicolumn{4}{c}{[1500 - 2000]/[5000 - 15000]/[15000 - 30000] (MIPS)} \\
        \midrule
        \multirow{1}{*}{MC of CN in Tier-1/2/3} & \multicolumn{4}{c}{[4 - 8]/[8 - 16]/[32 - 64]  (GB)} \\
        \midrule
        \multirow{1}{*}{DC of CN in Tier-1/2/3} & \multicolumn{4}{c}{[8 - 32]/[32 - 128]/[128 - 256]  (GB)} \\
        \midrule
        \multirow{1}{*}{DS of SC} & \multicolumn{4}{c}{[500 - 800] (Mb)} \\
        \midrule
        \multirow{1}{*}{RS of CN/SC} & \multicolumn{4}{c}{[0.7 - 0.9]/[0.9 - 0.99]} \\
        \midrule
        \multirow{1}{*}{Link BW} & \multicolumn{4}{c}{[100 - 500] (Mbps)} \\
        \midrule
        \multirow{1}{*}{Link RTT} & \multicolumn{4}{c}{[500 - 1200] (ms)} \\

        \bottomrule
    \end{tabular}
\end{table*}

\subsection{Other Scheduling Algorithms}
To gauge the efficiency of MOGA and compare its results, we also designed and implemented other heuristic solvers; namely, (1) Task Continuation Affinity (TCA), (2) Least Required CPU (LRC), (3) Most Data Size (MDS), (4) Most Reliability (MR), (5) Most Powerful (MP), and (6) Least Powerful (LP). TCA executes the first version of the service component on the user nodes if sufficient resources are available. In the cases of inadequate resources, it attempts with subsequent versions. If there are still insufficient resources after checking all versions, the service component is placed to the upper tier. Similarly, if the computing nodes in each tier do not have sufficient resources after checking all versions, the remaining service components are placed to the upper tier. LRC operates similarly to TCA, but only selects the version of service components that demand the least CPU for completion. MDS also operates similarly to TCA, but prioritizes running service components with the largest data size on user nodes or closer tiers to the users as long as resources are available. MR runs the version of the service components with the highest reliability on the computing node that has the highest reliability in the infrastructure. MP runs the least computationally demanding version on the most powerful node in terms of computational capacity. LP is the opposite of MP, where the most computationally demanding version is executed on the least powerful node.

\subsection{Evaluation Scenarios}
To evaluate the performance of MOGA, we consider four evaluation scenarios (i.e., problem instances): small-scale, medium-scale, large-scale, and xLarge-scale scenarios. Table \ref{table:scenarios} shows the specifications of all scenarios along with the characteristics of their assumes infrastructure and services. The values for the infrastructure reflect the scale (properties of all nodes and networking links), and thus selected carefully (based on our discussion with our industry partners) to reflect realistic infrastructures. To pressure all algorithms and push them to their limits (in terms of the quality of the solutions and the convergence time), our simulator then generates the service components and their requirements so that the total resource capacity of the service components constitutes 60-70\% of the total resource capacity of the infrastructure. Additionally, we assume that each helper can provide assistance to more than one user; that is; a helper can be part of several user-helper pairs.

\subsection{Configuration setup of MOGA}
The configuration of MOGA significantly affects its performance. We designed a grid-based tuning strategy to find the best configuration for our MOGA. To this end, we first ran the grid-based tuning for all scenarios and investigated the impacts of the parameters on the performance of MOGA in each scenario. 

\begin{figure*}[t]
    \centering
    \includegraphics[width=1\linewidth]{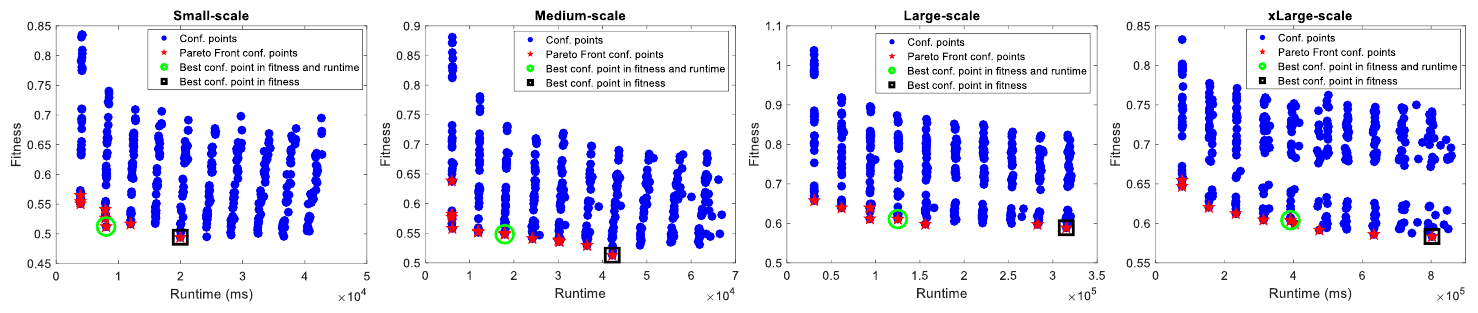}
    \captionsetup{
        justification=centering,
        font={small,stretch=0.9}
    }
    \caption{Selecting the best configuration for MOGA in different scales by the Pareto Front-based method}
    \label{fig:configurationSelection}
\end{figure*}

\begin{figure*}[t]
    \centering
    \includegraphics[width=1\linewidth]{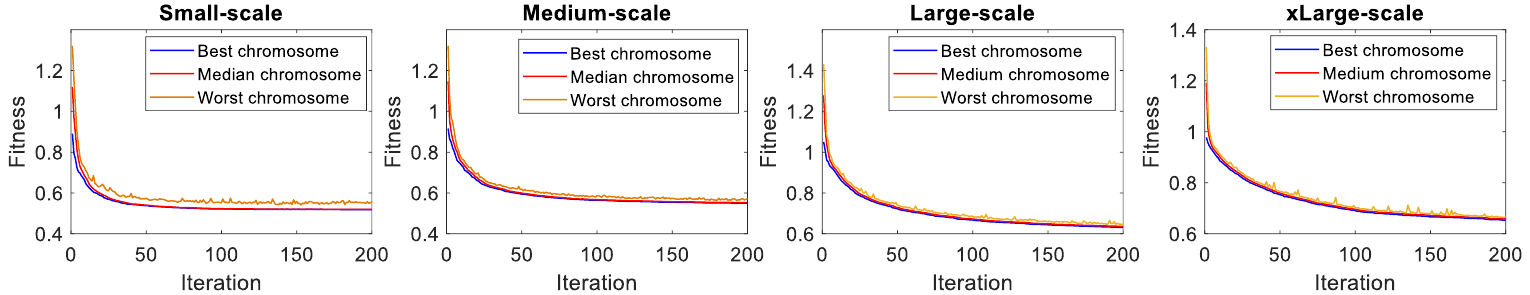}
    \captionsetup{
        justification=centering,
        font={small,stretch=0.9}
    }
    \caption{Convergence process of MOGA in different scales}
    \label{fig:convergence}
\end{figure*}

In MOGA, similar to any other GA-based solver, the fitness value tends to improve as the population size increases, and larger population sizes (almost) always yield better solutions.
However, running MOGA for large and extra-large (xLarge) scenarios with large population sizes led to unacceptable (more than 24 hours) convergence times. Our grid-based tuning procedure is to strike a balance between MOGA's quality of solutions and its convergence time. This is because some solutions (fitness points) produced through smaller population sizes may not differ much from better solutions found through larger population sizes at the cost of much higher convergence time. This implies that acceptable (not so optimal solutions) can be achieved with smaller population sizes, leading to a significant decrease in MOGA's runtime. To this end, we implemented an algorithm based on the Pareto Front method to select the configuration that strikes a balance between the `fitness value of the best solution' and the `MOGA's runtime'.

To start the algorithm, MOGA is run for every combination of `population-size', `elitism-rate', `cross-over rate', and `mutation rate' for a very limited number of scenarios. After each run, (1) the fitness value of the best solution and (2) MOGA's runtime are recorded for further processing. Figure \ref{fig:configurationSelection} shows sample records after collecting all points for different sizes of problem instances. Using these points, we aim to find several formulas to relate each MOGA's parameter to the characteristics of a problem instance; for example, finding a formula to select a reasonable population size for MOGA based on the number of user-helpers and total number of computing nodes in an infrastructure. Using such formulas to set MOGA's input parameters, we can then ensure that MOGA will (most likely) find the best solution (or one very close to it) within a reasonable amount of time.

To find a point that strikes a balance between the solution quality and MOGA's runtime, we first find the Pareto Front (curve) of these points. To that end, a configuration (out of all available points, shown in blue in Figure \ref{fig:configurationSelection}) is randomly selected, and all other configurations whose fitness and runtime are worse than the selected point are deleted. This process is repeated until no further points can be removed from the list. The calculated Pareto-front that represent `best configurations' (because no point is strictly better than any other point on the same Pareto front) are then sorted based on their population size, and their median is identified as the optimal population size for that scenario. Similarly, the same `best configurations' are then re-sorted based on their crossover rate, and their median is identified as the optimal crossover rate for that scenario. The same procedure (re-sorting) is repeated to select the best elitism and mutation rates.

Figure \ref{fig:configurationSelection} illustrates how our Pareto Front-based approach selects the best configuration to balance fitness and run-time on various scales. Based on the results in Figure \ref{fig:configurationSelection}, we observe that although the fitness value of the best configuration points in terms of fitness are relatively superior compared to the best configuration points selected by the Pareto Front-based method, these superiorities are negligible compared to the variations in runtime. Specifically, the optimal configuration point selected by the Pareto Front-based method shows 63\%, 57\%, 60\%, and 53\% better runtime in small-scale, medium-scale, large-scale and xLarge-scale scenarios, respectively. On the contrary, the best fitness configuration points only show improvements of 6\%, 4\%, 3\%, and 7\% compared to Pareto Front-based configuration points on all scales, respectively.

The optimal configurations considered by our Pareto Front-based method for all scenarios are indicated in Table \ref{table:parameters}. As shown in Table \ref{table:parameters}, the optimal population size increases slightly with the problem size. On the contrary, the optimal value of the mutation rate and the selection size are considered 1\% and 10\% for all scenarios, respectively. Furthermore, the crossover rate falls within the range 60\% to 80\%, which means that a larger problem size requires a relatively higher crossover rate. Using the information in Table \ref{table:parameters}, we can provide an equation to estimate the optimal value for population size. Equation \eqref{eq:27} estimates the optimal population size in various problem sizes, where $(X\times Y)$ is the chromosome length, where $N$, $M$, and $K$ represent the total number of user nodes, helper nodes, and computing nodes, respectively. $\alpha$, $\beta$, and $\gamma$ are fixed values that we calculated (using the curve fitting algorithm) $\alpha = 0.9$, $\beta = 0.16$ and $\gamma = 0.16$ to estimate the optimal population size. 

\begin{equation}\label{eq:27}
ps \approx 100^{\alpha} \cdot (X\times Y)^{\beta} \cdot (N+M+K)^{\gamma} 
\end{equation}

\begin{table}
    \caption{Optimal values for MOGA parameters}
    \label{table:parameters}
    \renewcommand{\arraystretch}{1}
    \small
    \begin{tabular}{m{1.2cm} m{1cm} m{1cm} m{1cm} m{1cm}}
        \toprule
        Parameter & Small-scale & Medium-scale & Large-scale & xLarge-scale \\
        \midrule
        \midrule
        $ps$ & 200 & 300 &  400 & 500 \\
        \midrule
        $cr$ & 60\% & 70\% &  70\% & 80\% \\
        \midrule
        $mr$ & 1\% & 1\% &  1\% & 1\% \\
        \midrule
        $ss$ & 20 & 30 &  40 & 50 \\
        \midrule
        $it$ & 50 & 100 &  150 & 200 \\
        \bottomrule
    \end{tabular}
\end{table}

Similarly, the crossover rate ($cr$) can be estimated using Equation \eqref{eq:28}, where $c$ is calculated using Equation \eqref{eq:29} where $\delta = 0.0003$ and $\varepsilon = 0.04$. Since we identified an optimal crossover rate 60\% to 80\% through our fine-tuning process for MOGA, in Equation \eqref{eq:28}, we consider specific conditions to set the crossover rate within the range 60\% to 80\%.

\begin{equation}\label{eq:28}
cr \approx \begin{cases}c & c \leqslant 0.8\\0.8 & c > 0.8\end{cases}\\
\end{equation}

\begin{equation}\label{eq:29}
c = 0.6+(X\times Y)\cdot\delta + (N+M+K)^\varepsilon 
\end{equation}

Figure \ref{fig:convergence} shows the MOGA convergence process in terms of the best, median, and worst chromosomes of the population in the different scenarios. It is evident that MOGA converges rapidly on the small and medium scale, particularly within the first 100 iterations. Therefore, we set the number of MOGA iterations to 50 and 100 for small and medium scales to achieve better run-time efficiency. We also set the number of MOGA iterations at 150 and 200 for the large and xLarge scales, respectively, because no notable improvements in the fitness value were observed across these scenarios after 150 and 200 iterations. As a result, as the problem size increases, MOGA requires more iterations. We also developed Equation \eqref{eq:30} to estimate sufficient number of iterations for MOGA, where $\zeta$ = 0.9, $\eta$ = 0.69, and $\theta$ = 0.1. 

\begin{equation}\label{eq:30}
it \approx 50^\zeta+(X\times Y)^\eta \cdot (N+M+K)^\theta
\end{equation}

Figure \ref{fig:convergence} provides more information on the overall performance of the MOGA. We investigated the median and worst chromosomes to assess the performance of MOGA in terms of exploration and exploitation. The results show that the median chromosomes consistently follow the best chromosomes, indicating that MOGA effectively exploits good solutions. The worst chromosomes also show an improvement over time, indicating that the algorithm maintains diversity, while still exploring different regions of the solution space.

\subsection{Response time evaluation}
\begin{figure*}[t]
    \centering
    \includegraphics[width=1\linewidth]{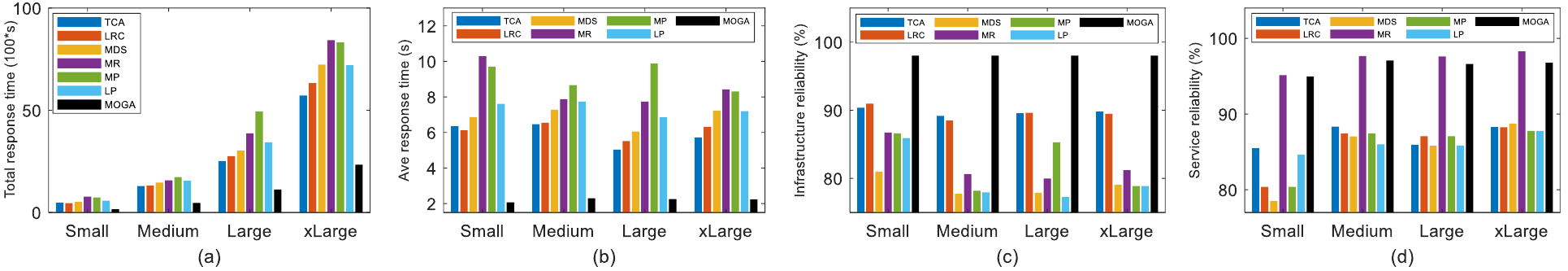}
    \captionsetup{
        justification=centering,
        font={small,stretch=0.9} 
    }
    \caption{Evaluations based on response time and system reliability}
    \label{fig:resutls}
\end{figure*}

Figure \ref{fig:resutls} (a) compares the total response time achieved by MOGA with the heuristic algorithms in our four scenarios. More specifically, MOGA leads to an average improvement of 71\%, 67\%, 65\%, and 66\% in total response time across small-scale, medium-scale, large-scale, and xLarge-scale, respectively. Furthermore, Figure \ref{fig:resutls} (b) compares the average response time of the algorithms for each service component. As depicted in Figure \ref{fig:resutls} (b), the average response time of MOGA for a service component is approximately 2 seconds on the different scales, while the heuristic algorithms indicate average response times above 5 seconds. The obtained results related to response time show the superior performance of our MOGA compared to the heuristic algorithms and indicate MOGA's capability (as a meta-heuristic algorithm) to identify and implement efficient solutions with minimal response times.

\subsection{Infrastructure and service reliability evaluation}
Figure \ref{fig:resutls} (c) illustrates a comparison of the infrastructure reliability achieved by both MOGA and the heuristic algorithms. On average, MOGA shows an infrastructure reliability close to 98\%, whereas the reliability achieved by other heuristic algorithms falls below 91\%. Furthermore, Figure \ref{fig:resutls} (d) compares the service reliability between MOGA and the heuristic algorithms. The results in Figure \ref{fig:resutls} (d) indicate that MOGA consistently maintains a high level of service reliability in all scenarios, with an average of 96\%. In addition, a high level of service reliability is observed for MR in all scenarios, with an average of 97\%. As mentioned above, MR prioritizes the placement of service components with higher reliability scores, resulting in high service reliability results. However, MR faces limitations in providing desirable infrastructure reliability and response time. Regarding the other heuristic algorithms, they provide a service reliability ranging from 75\% to 88\%.

Based on the findings presented in this subsection, MOGA provides a high degree of infrastructure reliability and an excellent degree of service reliability. This maximization in reliability, both in hardware and software aspects, is vital in minimizing interruptions arising from reliability-related issues within edge-to-cloud environments, mainly when it involves the demanding requirements of AR/VR applications. The results clearly illustrate the scalability of MOGA that consistently maintains a high degree of infrastructure and service reliability on various scales of the system. It indicates its adaptability and effectiveness regardless of the scale of the system.

\subsection{Algorithm runtime evaluation}
Table \ref{table:runtime} presents the runtime of both MOGA and the heuristic algorithms. As expected, the run-time of MOGA is higher compared to the heuristic algorithms due to the nature of its operators. However, we made a significant reduction in MOGA runtime by optimizing its configurations.

\begin{table}
    \caption{Runtime of algorithms (s)}
    \label{table:runtime}
    \renewcommand{\arraystretch}{1}
    \small
    \begin{tabular}{m{1.2cm} m{1cm} m{1cm} m{1cm} m{1cm}}
        \toprule
        Algorithm & Small-scale & Medium-scale & Large-scale & xLarge-scale \\
        \midrule
        \midrule
        TCA & 0.0026 & 0.0109 & 0.0411 & 0.0499 \\
        \midrule
        LRC & 0.0014 & 0.0026 & 0.0087 & 0.0239 \\
        \midrule
        MDS & 0.0021 & 0.0317 & 0.0464 & 0.0622  \\
        \midrule
        MR & 0.0015 & 0.0328 & 0.0299 & 0.0899 \\
        \midrule
        MP & 0.0024 & 0.0058 & 0.0301 & 0.0951 \\
        \midrule
        LP & 0.0013 & 0.0351 & 0.084 & 0.1075 \\
        \midrule
        MOGA & 8.73 & 79.74 & 530.78 & 2678.25 \\
        \bottomrule
    \end{tabular}
\end{table}

\begin{figure*}[t]
    \centering
    \includegraphics[width=1\linewidth]{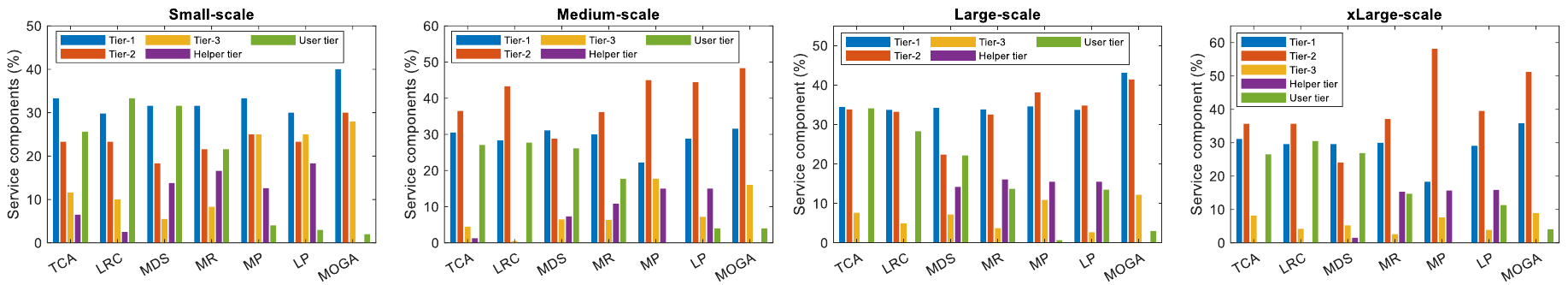}
    \captionsetup{
        justification=centering,
        font={small,stretch=0.9} 
    }
    \caption{Distribution of service components over the infrastructure in different scales}
    \label{fig:percentage}
\end{figure*}

\subsection{Distribution of service components over the infrastructure}
Figure \ref{fig:percentage} depicts the percentage of distribution of service components using the algorithms and provides detailed information on the operation of the algorithms.

Based on the results of Figure \ref{fig:percentage}, it is clear that MOGA never assigns service components for execution on helper nodes. The reason is that the helper nodes communicate with the users through the cloud layer. Therefore, it is rational to place service components on superior nodes in proximity to user nodes rather than helper nodes. Additionally, MOGA places only a minor portion of the service components to user nodes. In fact, MOGA minimizes the usage of user nodes due to their constrained computational and storage capacities. Instead, MOGA prefers to employ more robust computing nodes in Tier-1 and Tier-2, which are located near user nodes, to improve performance. Therefore, we can conclude that MOGA minimizes response time by decreasing the placement of service components in Tier-3 and the rare usage of helper and user nodes. This not only leads to a decrease in the overall transmission time but also reduces execution time with respect to leveraging more powerful computing nodes at the network's edge. However, it is essential to note that MOGA also optimally selects the appropriate version of the service component on the appropriate computing nodes in each tier, leading to significant improvements both in response time and system reliability. Although the selection of nodes in Tiers 2 and 3 seems trivial, and thus may falsely give the belief that the cost of running MOGA (or other meta-heuristics for that matter) is not worth, it is essential to note that other heuristics that mentioned in this paper and roughly deploy such simplistic procedures could not lead to MOGA's high quality answers. This is mainly due to the fact that simultaneously selecting both computing nodes and software versions adds extra layers of complexity that simple heuristics cannot tolerate. 

\section{\uppercase{Conclusion}}
\label{sec:Conclusion}
In this paper, we introduce a MOGA for optimal service placement in edge-to-cloud AR/VR systems. The primary objectives were to minimize service response time, maximize infrastructure reliability, and achieve the highest service reliability by optimally placing service components on computing nodes, user nodes, and helper nodes in the edge-to-cloud infrastructure. We devised a robust fine-tuning strategy to attain optimal configurations for our MOGA in order to strike a balance between the MOGA's runtime and the quality of its solutions. We also implemented a simulator to validate the proposed MOGA's effectiveness. Through extensive simulations and measurements on various scales, we showed the importance of designing MOGA (as well as other meta-heuristic) to simultaneously optimize for response time and infrastructure and service reliability. We also assessed the performance of MOGA in terms of the distribution of service components over the infrastructure, illustrating how our MOGA provides an optimal solution for the placement of AR/VR services compared to the other rather simple heuristic algorithms.

\section*{\uppercase{Acknowledgements}}
\small
Parts of this work have been supported by the Knowledge Foundation of Sweden (KKS).
\bibliographystyle{apacite}


\end{document}